\documentclass{PoS}

\usepackage{amsmath}

\title{Matrix Models of Fuzzy Field Theories}

\ShortTitle{Matrix Models of Fuzzy Field Theories}

\author{M\'aria \v Subjakov\'a, \speaker{Juraj Tekel}\\
        Department of Theoretical Physics and Didactics of Physics\\
        Faculty of mathematics, physics and informatics\\
        Comenius University, Bratislava, Slovakia\\
        E-mail: \email{subjakova@fmph.uniba.sk}, \email{tekel@fmph.uniba.sk}}

\abstract{We briefly review the connection between the fuzzy field theories and matrix models and describe the main features of the models that appear. We summarize the different approaches to their analysis, some of the recent results and the challenges to be addressed in the future.}

\FullConference{Corfu Summer Institute 2017 "School and Workshops on Elementary Particle Physics and Gravity"\\
                 2-28 September 2017\\
                 Corfu, Greece}

\def\ep{\varepsilon}

\def\half{\frac{1}{2}}

\def\K{\mathcal{K}}

\def\lr#1{\left(#1\right)}

\def\tr#1{\textrm{Tr}\Big(#1\Big)}
\def\trl#1{\textrm{Tr}\lr{#1}}

\begin{document}

\section{Introduction}

Commutative limits of naive noncommutative field theories are very different from their commutative counterparts. A noncommutative space tends to its commutative version in the limit of the vanishing noncommutativity parameter, but the limit of the field theory does remember its origin. This is a result of the UV/IR-mixing phenomenon. Due to the minimal length, quanta can not be compressed into arbitrarily small volumes. Short distance in one direction becomes long distance in the other directions and the separation between UV and IR degrees of freedom is lost.

As a result, one obtains nonlocal terms in the effective action of the theory, which do not vanish in the commutative limit.

One of the consequences of this effect is the existence of a purely noncommutative phase of scalar field theories. In this phase the field does not oscillate around one given value in the whole space, but rather forms stripes of oscillations around different vacua of the theory, and the translational symmetry of the underlying space is spontaneously broken. In this short review, we discuss a description of this phenomenon for compact noncommutative spaces, or fuzzy spaces, by the means of a certain class of matrix models.

We first very briefly review the construction of fuzzy spaces and fuzzy field theories and numerical results for their spontaneous symmetry breaking patterns. We then describe how one can treat such field theories by considering certain hermitian matrix models and introduce the main ingredient, the contribution to the probability distribution due to the kinetic term of the field theory. We review previous approaches to treatment of such contribution and before concluding with the most interesting questions to be answered in the future, we present two models that approximate the kinetic term in a new way.

\section{Fuzzy field theory}

Noncommutative spaces provide a setting with a shortest possible distance without any loss of the underlying symmetry. Fuzzy spaces are a finite volume example, where we have a finite number $N$ of ''Planck cells'' of the space. There are however no sharp boundaries between them and the whole space is blurred, without a good definition of a point.

The hallmark example of this is the fuzzy sphere \cite{hoppe,sF21,sF22}. Its construction can be carried out as follows. The information about the commutative sphere is encoded in the algebra of functions generated by the elements
\begin{equation}\label{gen_com}
x_i x_i = R^2 \ , \ x_i x_j - x_j x_i=0\ .
\end{equation}
With a pointwise multiplication of functions, this algebra is commutative by definition.

To obtain the fuzzy sphere, we deform the conditions (\ref{gen_com}) to the form
\begin{equation}\label{gen_noncom}
x_i x_i = \rho^2 \ , \ x_i x_j - x_j x_i=i \theta \ep_{ijk} x_k\ ,
\end{equation}
where the right hand side of the condition is dictated by the anticipated rotational symmetry. These conditions can be satisfied as an ${N=2j+1}$ dimensional representation of $SU(2)$, namely
\begin{equation}\label{su2_realization}
x_i=\frac{2R}{\sqrt{N^2-1}} L_i\ , \ \theta=\frac{2R}{\sqrt{N^2-1}} \ , \ \rho^2=\frac{4R^2}{N^2-1} j(j+1)=R^2\ .
\end{equation}
The algebra of functions on the fuzzy sphere is then the algebra generated by these generators, i.e. the algebra of the ${N\times N}$ matrices. Other fuzzy spaces are defined in a similar fashion \cite{stenposi}.

The field theory on the fuzzy space is then defined recalling the commutative case, where the action for an euclidean real scalar field is given by
\begin{equation}\label{com_action}
S({\color{black}\Phi})=
	{\color{black}\int dx\bigg[}
	\half
	{\color{black}\Phi} {\color{black}\Delta} {\color{black}\Phi}
	+\half m^2 {\color{black}\Phi}^2
	+V({\color{black}\Phi}){\color{black}\bigg]}\ .
\end{equation}
The dynamics of the theory is given by correlation functions of the field in terms of a functional integral
\begin{equation}\label{com_correlation}
	\left\langle F\right\rangle=\frac{\int {\color{black}d\Phi}\, F({\color{black}\Phi})e^{-S({\color{black}\Phi})}}{\int {\color{black}d\Phi}\,e^{-S({\color{black}\Phi})}}\ .
\end{equation}

We need to establish a dictionary between the commutative and the fuzzy cases \cite{stenposi}. As we have seen the field is given by an ${N\times N}$ matrix. The functional integral is then given by an integral over the space of all hermitian matrices with the measure
\begin{equation}
dM=\prod_i dM_{ii}\prod_{i<j}d\textrm{Re}M_{ij}\,d\textrm{Im}M_{ij}\ .
\end{equation}
The spacetime integral becomes a properly normalized trace, since it is the sum of the values of the field. The derivative, as a change with respect to a small transformation, becomes the commutator with the transformation generator. Translating all the objects in (\ref{com_action},\ref{com_correlation}) we obtain
\begin{align}
S({\color{black}M})\,=\,&
	{\color{black}\frac{4\pi R^2}{N}\textrm{Tr}\bigg[}
	\half
	{\color{black}M}{\color{black}\frac{1}{R^2} [L_i,[L_i,}{\color{black}M}{\color{black}]]}
	+\half m^2 {\color{black}M}^2
	+V({\color{black}M}){\color{black}\bigg]}\ ,\label{noncom_action}\\
\left\langle F\right\rangle\,=\,&\frac{\int {\color{black}dM}\, F({\color{black}M})e^{-S({\color{black}M})}}{\int {\color{black}dM}\,e^{-S({\color{black}M})}}\ .\label{noncom_correlation}
\end{align}
As before, the answers to the field theory questions are hidden in the relevant correlators \cite{bal,szabo}. Without too much surprise the noncommutative theory is different from the commutative one. It however turns out that it is very different, as this difference even survives the commutative limit. Note, that the large $N$ limit in (\ref{su2_realization}) leads to $\theta\to0$, which turns (\ref{gen_noncom}) into (\ref{gen_com}). In the limit of a large number of cells, the fuzzy sphere becomes an ordinary sphere. But the field theory (\ref{noncom_action},\ref{noncom_correlation}) does not become its commutative counterpart. There are effects that are carried over from the noncommutative case, due to the phenomenon of UV/IR-mixing \cite{uvir1,uvir2}.

For our purposes, the interesting difference is in the symmetry breaking patterns of the theory. For example the commutative $\Phi^4$ scalar field theory has two different phases \cite{comr1,comr3,comr1num,comr2num}. The disorder phase, where the field oscillates around the value $\Phi=0$ and the uniform order phase, where the field oscillates around one of the global minima of the potential. Existence of a third phase has been estabilished computationaly \cite{NCphase1,noncomphase3} and observed numerically\footnote{See \cite{ccc} for a review.} for the fuzzy sphere \cite{nummartin,denjoenum1,paneronum,denjoenum2,num14,samo}, fuzzy sphere with a commutative time \cite{num_RSF2}, fuzzy disc \cite{num_disc} and noncommutative plane \cite{num14panero2}. This phase, a non-uniform order or a striped phase, breaks the translational symmetry of the space and the field does not oscillate around one given value of the field in the whole space. And most importantly it remains present even in the large $N$ limit of the theory.

In the rest of the paper we present matrix models, which describe, qualitatively and to some extent also quantitatively this interesting phase structure of fuzzy field theories.

\section{Matrix models of fuzzy field theory}

Without much explanation needed, the fuzzy field theory (\ref{noncom_correlation}) is a random matrix model, since the interesting quantities are given as expectation values in a random matrix model with the probability distribution
\begin{equation}
P(M)dM=e^{-S(M)}dM
\end{equation}
and the field theory action (\ref{noncom_action}). As mentioned in the introduction, models without the kinetic term are well known and we shall give an extremely brief overview their treatment before we proceed to the full model. For concreteness we will deal with the $\Phi^4$ theory from now on.

\subsection{Pure matrix models}

The relevant matrix model without the kinetic term is simply
\begin{equation}\label{quartic_model}
S(M)=\half r \trl{M^2}+g\trl{M^4}
\end{equation}
and the key insight is that such probability density is invariant under the conjugation of the matrix $M$ with a unitary matrix ${U\in SU(N)}$. This allows us to diagonalize the matrix $M\to U\Lambda U^\dagger$, with $\Lambda=diag(\lambda_1,\ldots,\lambda_N)$ the diagonal matrix of the eigenvalues of $M$. The integration measure in the terms of the eigenvalues and the angular degrees of freedom changes in the following way
\begin{equation}
dM=dU \lr{\prod_{i=1}^N d\lambda_i}\times \prod_{i<j}(\lambda_i-\lambda_j)^2\ .
\end{equation}
The extra jacobian can be exponentiated to introduce a logarithmic repulsion between pairs of eigenvalues. There is nothing that depends on $U$ in the probability measure and the $dU$ integral is trivial if we ask $U$-invariant questions. The problem becomes purely eigenvalue one and can be solved in the large $N$ limit by the saddle point approximation \cite{brezin,mmodels,JT15APS}.

\begin{figure}
\begin{center}
\includegraphics[width=.3\textwidth]{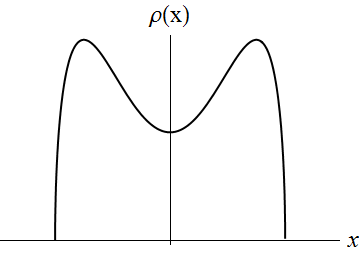}\ 
\includegraphics[width=.3\textwidth]{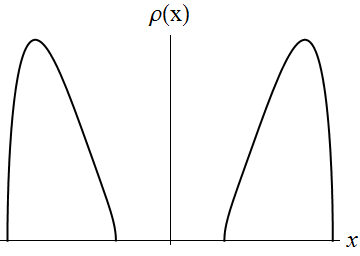}
\end{center}
\caption{The two possible types of the eigenvalue distribution $\rho(x)$ of the model (\ref{quartic_model}). The gap opens at $r=-4\sqrt g$. The graphs have different scales.}
\label{fig1}
\end{figure}

The key feature of such model is the phase transition. For values ${r>-4\sqrt{g}}$ the eigenvalue distribution $\rho(x)$ remains supported on a single symmetric interval. But for values ${r<-4\sqrt{g}}$ the eigenvalues split, as the two wells of the potential are too deep, and $\rho(x)$ is supported on two disjoint intervals (figure \ref{fig1}).

As we will see in the next sections, a correction to (\ref{quartic_model}) can lead to a situation, where a stable distribution which lives completely in one of the wells of the potential can become possible.

\subsection{Kinetic term of the field theory}

The complete matrix model of the $\Phi^4$ theory is given by the action (\ref{noncom_action}), which we rewrite as
\begin{equation}
S(M)=\half\trl{M\K M}+\half r\, \trl{M^2}+g\,\trl{M^4}
\end{equation}
by rescaling the parameters and the matrix and by introducing
\begin{equation}
\K M = [L_i,[L_i, M]]\ .
\end{equation}

The kinetic term of the action is not invariant under conjugation with a unitary matrix and it does have a very non-trivial $U$-dependence. Therefore in this case the integration over the angular degrees of freedom cannot be straightforwardly  performed.

\section{Kinetic term effective action}

However, since the angular integration is a function of the matrix eigenvalues only, we can lift it into the exponential. We define the kinetic term effective action \cite{steinacker05,nonperturbative}
\begin{equation}\label{kin_term}
e^{-S_{eff}}=  \int dU e^{-\half \trl{M\mathcal{K} M} }
\end{equation}
and consider the following probability distribution for the eigenvalues $\lambda_i$ only
\begin{equation}
S(M)=S_{eff}(\Lambda)+\half r\, \sum_i \lambda_i^2+g\,\sum_i \lambda_i^4-2\sum_{i< j}\log|\lambda_i-\lambda_j|\ .
\end{equation}

The key goal in the analytical treatment of the fuzzy scalar theory is to determine the eigenvalue dependence of the effective action.

\subsection{Perturbative calculations}

Perturbative treatment of the angular integral was first introduced in \cite{perturbative_1}, where the perturbative expansion of the integral in terms of kinetic action was made
\begin{equation}
\int dU e^{-\half \trl{M\mathcal{K} M} }= \sum_n \frac{1}{n!} \int dU \lr{-\half\trl{M\mathcal{K}M}}^n = \sum_n I_n
\end{equation}
and the first few integrals were computed, in the spirit of the character expansion method for angular integrals. This method was further extended in \cite{perturbative_2} where the integrals $I_n$ were calculated up to the third order using several group techniques and identities. 

In \cite{perturbative_bootstrap}, a different method to obtain the same  expansion was used. The  symmetries of the kinetic term were considered
\begin{equation}
De^{-S_{eff}}= O(\Lambda)e^{-S{eff}}\ .
\end{equation}
with $D$ being some differential operator and $O(\Lambda)$ a fuction of the eigenvalues only. Such symmetries were then used  to put constraints on the expansion coefficients resulting in the following perturbative expansion of the effective action up to the fourth order
\begin{align}
S_{eff}(M)=&\half \bigg( \half t_2-\frac{1}{24}t_2^2+\frac{1}{2880}t_2^4\bigg)- \frac{1}{432}t_3^2-  \frac{1}{3456}\bigg(t_4-2t_2^2\bigg)^2 + \ldots \ ,\label{multitrace}
\end{align}
in term of the (normalized) symmetrized moments
\begin{equation}\label{sym_moments}
t_n = \frac{1}{N} \tr{M-\frac{1}{N}\trl{M}}^n. 
\end{equation}
It is possible to study this multitrace model, look for stable asymmetric solutions and for the features of the phase diagram. It has been done, both numerically and analytically \cite{JT15APS,JT15,bbb}, and it has been shown that the results are not satisfactory. The model does not have a triple point and behaves very differently from expectations close to the origin of the parameter space.

\subsection{Non-perturbative bootstrap}\label{sec42}

In \cite{nonperturbative}, another treatment of the angular integral (\ref{kin_term}) was developed. Using this method it is possible to determine certain part of the effective action analytically but it has little to say about the rest of it.  

The key observation is that in the case of ${g=0}$,  the matrix model corresponding to $\Phi^4$ is solvable analytically in the commutative limit \cite{steinacker05,NPT12}. In such free theory, introducing the kinetic term into the action only rescales the probability distribution of the eigenvalues, but the distribution remains of the same shape. This effect can be reproduced by the effective action of the form
\begin{equation}
S_{eff} = \half F_2(t_2)+ \mathcal{R}\ ,
\end{equation}
where $F_2(t_2)$ is the part of the action that causes rescaling of the probability distribution and it is equal to\footnote{Note, that the small $t_2$ expansion correctly reproduces (\ref{multitrace}).}
\begin{equation}\label{F2}
F_2(t_2) = \log \bigg ( \frac{t_2}{1-e^{-t_2}} \bigg).
\end{equation}
The remainder $ \mathcal{R}$ gives a vanishing contribution for ${g=0}$, i.e. when evaluated using (\ref{sym_moments}) of the Wigner semicircle distribution, which is the solution in the free case. Therefore $ \mathcal{R}$  must contain only the products of at least two such terms, which are simply $t_{2n-1}$ or $\big(t_{2n}- \frac{(2n)!}{n!(n+1)!}t_2\big)$.

\begin{figure}
\begin{center}
\includegraphics[width=.3\textwidth]{1cut.png}\ 
\includegraphics[width=.3\textwidth]{2cut.png}\ 
\includegraphics[width=.3\textwidth]{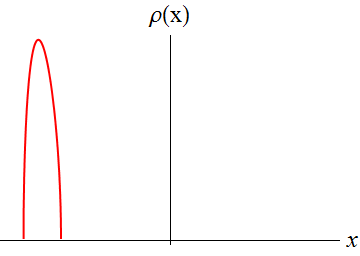}
\end{center}
\caption{The three possible types of the eigenvalue distribution $\rho(x)$ of the model (\ref{F2}). The graphs have different scales.}
\label{fig2}
\end{figure}

The remainder term can be dropped as an approximation and the resulting model studied \cite{JT15APS,JT18}. There is a region in the parameter space where the attraction among the eigenvalues introduced by $F_2$ overcomes the repulsion and an asymmetric solution living in only one of the wells of the potential (figure \ref{fig2}) remains stable and the preferred solution to the model.
The diagram obtained by this approximation reproduces the features of the diagram obtained numerically qualitatively and to certain extent also quantitatively \cite{JT18}. 

\subsection{Beyond the second moment}

As it was mentioned in the preceding sections the perturbative expansion is not a very successful tool to study the structure of the phase diagram of the $\Phi^4$ theory. It seems that to reproduce the phase structure obtained through the numerical simulations, one needs to arrange the known terms of the expansion into some non-perturbative package. It was shown that some progress can be made considering terms involving only the second moment of the distribution.
 
One way of proceeding further in the analytical treatment of the theory could thus be to arrange also the known $\mathcal{R}$ terms into packages similar to (\ref{F2})   
\begin{equation}\label{model1}
S_{eff} = \half F(t_2) + F_3(t_3) + F_4 (t_4-2t_2^2)\ .
\end{equation}
In principle, the functions ${F_3(t)}$ and ${F_4 (t)}$ can be chosen arbitrarily, but must have the correct perturbative expansion of the effective action (\ref{multitrace}). However, it is reasonable to demand that these functions satisfy few conditions.

In order to reproduce the phase diagram of the $\Phi^4$ theory obtained in numerical simulations, we want the phase transition line between the disorder and the non-uniform order phase going through the origin of the parameter space. This leads to the following constraint on the behaviour of the functions ${F_3(t)}$ and ${F_4 (t)}$

\begin{equation}
F_3'(t), F_4' (t) \to 0 \ \textrm{as}\  |t| \rightarrow \infty\ .
\end{equation}
Thus, this functions cannot increase in a rate  higher than $\log|t|$ as $|t| \rightarrow \infty$.

We also expect the higher terms in the perturbative expansion to give a significant contribution and therefore not be too small in comparison with first terms. The good trial functions could be for example:
\begin{displaymath}
 -A\log(1+t^2)\ ,\ A\lr{\frac{1}{1+t^2}-1}\ .
\end{displaymath}
It would  be interesting to study such matrix models and analyze phase structure sensitivity  on the particular choice of the functions $ F_3(t)$ and $F_4 (t)$.

\subsection{Two body interaction}

In this section, we present a different method which might bring new insight to the problem of the structure of the effective action. We restrict our attention only to the double-trace terms of the multitrace expansion (\ref{multitrace}) and try to find a function which correctly reproduces this part of the expansion in the form of a two body interaction
\begin{align}
S_{eff} & =\frac{1}{4} c_2-\frac{1}{4}c_1^2-\frac{1}{24}c_2^2-\frac{1}{432}c_3^2-\frac{1}{3456}c_4^2+\ldots\ , \label{two_body}\\
S_{eff} & = \frac{1}{4} c_2+ \sum_i \sum_j f(\lambda_i,\lambda_i)+\ldots\ ,
\end{align}
where $c_n = \trl {M^n}/N$. The main motivation for this is the fact, that the analysis of multritrace models of the previous sections becomes technically very involved and we should look for a different kind of expansion of the effective action.

Interestingly, by choosing the  function $f(\lambda_i,\lambda_j)$ of the form
\begin{equation}
f(\lambda_i,\lambda_j)= a\log\lr{1-b\,\lambda_i \lambda_j}
\end{equation}
with parameters equal to $a=3/2$, $b=1/6$ one can correctly reproduce all four relevant coefficients in (\ref{two_body}). 
Investigating the matrix model with the effective action given by
\begin{equation}\label{model2}
S_{eff}= \frac{1}{4} c_2 + \frac{3}{2}\log\lr{1-\frac{1}{6}\lambda_i \lambda_j}
\end{equation}
could thus lead to some more information about the phase structure of $\Phi^4$ theory.

\section{Conclusions}

To conclude this short report let us mention several questions that are still unanswered.

Approach outlined in the section \ref{sec42} can be quite straightforwardly generalized to more complicated fuzzy spaces. For some, there are numerical and perturbative results to compare against \cite{num_RSF2,num_disc,num14panero2,samanfuzzydisc,rsfper}, for some, such as $\mathbb C P^n$ or the fuzzy four sphere, results would be an interesting prediction for the future numerical investigation. Pursuing this line of research further would be a very nice check of the consistency of the whole approach, suggested by the results for the fuzzy sphere.

The presence of the striped phase in the commutative limit of the phase diagram is related to the UV/IR-mixing. Modifications of the theories with no UV/IR-mixing are known for the fuzzy sphere \cite{nouvir1} and the noncommutative plane \cite{nouvir2,nouvir3}. It would be very interesting to investigate these theories, both analytically and numerically, and look for the presence of the striped phase.

And finally, the journey towards the computation of the angular integral (\ref{kin_term}) has barely started. The perturbative calculations become ridiculously complicated before giving enough insight into the structure of the kinetic term effective action and nonperturbative approaches give us information only about a small part of the effective action. At the moment, we are out of tricks and spells to battle the monster and  the models (\ref{model1}) and (\ref{model2}) are but a blind guess into the unknown we face.

\acknowledgments{
The stay of both M.\v S and J.T. at the \emph{Corfu Summer Institute 2017} has been supported by the COST Action MP1405 QSPACE. This work was supported by the VEGA 1/0985/16 grant and the work of J.T. is supported by the \emph{Alumni FMFI} foundation as a part of the \emph{N\'{a}vrat teoretikov} project.
}

\end{document}